\documentclass[aps,prb,twocolumn,superscriptaddress,showpacs]{revtex4}

\usepackage{amsmath}
\usepackage{amssymb}
\usepackage{graphicx}

\begin{document}

\title{Nonthermal broadening in the  conductance of
double quantum dot structures}

\author{L.~Oroszl\'any}
\affiliation{Department of Physics, Lancaster University,
Lancaster, LA1 4YB, UK}

\author{A.~Korm\'anyos}
\affiliation{Department of Physics, Lancaster University,
Lancaster, LA1 4YB, UK}

\author{J.~Koltai}
\affiliation{Department of Biological Physics, E{\"o}tv{\"o}s
University, H-1117 Budapest, P\'azm\'any P{\'e}ter s{\'e}t\'any 1/A,
Hungary}

\author{J.~Cserti}
\affiliation{Department of Physics of Complex Systems, E{\"o}tv{\"o}s
University, H-1117 Budapest, P\'azm\'any P{\'e}ter s{\'e}t\'any 1/A,
Hungary}

\author{C.~J. Lambert}
\affiliation{Department of Physics, Lancaster University,
Lancaster, LA1 4YB, UK}

\begin{abstract}
We study the transport properties of a double quantum dot (DQD)
molecule at zero and at finite temperature. The
properties of the zero temperature conductance  depends on whether
the level attraction between the symmetric and antisymmetric
states of the DQD, produced by the coupling to the leads, exceeds
or not the interdot tunneling. For finite temperature we find  a
remarkable nonthermal broadening effect of the conductance
resonance, when the energy levels of the individual dots are
detuned.

\end{abstract}


\maketitle

\section{Introduction}

Double quantum dot (DQD) systems exhibit a wide range of
interesting and fundamental physical phenomena, such as  Coulomb
blockade oscillations of conductance
\cite{ref:waugh-1,ref:waugh-2,ref:stafford,ref:matveev,ref:kotlyar,ref:nazarov,
ref:cheng, ref:vaart,ref:klimeck,ref:pohjola,ref:wiel}, the formation of
"double quantum dot molecule"
\cite{ref:blick-1,ref:blick-2,ref:stafford,ref:pohjola,ref:ziegler,ref:pioro,ref:wiel,
ref:huttel,ref:graeber,ref:hatano,ref:sapmaz,ref:ota,ref:pi,ref:biercuk,
ref:jorgensen,ref:fasth,ref:pfund}
or the Kondo effect\cite{ref:melloch-1,ref:melloch-2,ref:rushforth,ref:schroer}.
Experimentally,  DQDs have been formed
in semiconductor
heterostructures\cite{ref:waugh-1,ref:waugh-2,ref:blick-1,ref:blick-2,
ref:wiel,ref:pioro,ref:hatano, ref:huttel,ref:ota,ref:pi}, in single wall carbon
nanotubes\cite{ref:biercuk,ref:graeber,ref:sapmaz, ref:jorgensen}
and in InAs  nanowires\cite{ref:fasth,ref:pfund}. In recent years
interest in DQDs has also been   driven by the quest for a
solid-state based qubit, the elementary building block of a
quantum computer. In the presence of an interdot coupling $t_c$,
coherent electrons states can extend  over  the whole DQD system,
resembling therefore the formation of  chemical bonds in
molecules. This interdot coupling $t_c$ controls the exchange
interaction of electron spins, assumed to be localized in each of
the dots, and hence eventually the operation of a corresponding
solid-state qubit system\cite{ref:loss} as well.

Another interesting phenomenon, seen in experiments\cite{ref:wiel}
performed in Coulomb blockade regime at high source-drain voltages
is that the levels of one of the quantum dots (QDs) can act as a
low-temperature filter for the other QD. This means that on
detuning the energies of the levels participating in the resonant
tunnelling process, the width of the resonance peak in conductance
is independent of the temperature.

Surprisingly, a complete analytic description of non-thermal
broadening in not currently available, despite the fact that this
effect is alluded to in the early work of Ref.~\onlinecite{ref:nazarov} and
has been demonstrated numerically in Ref.~\onlinecite{ref:ziegler}
for coherently coupled quantum dots at large source-drain voltages
$e|V|\gg t_c$. A complete analytical theory of coherently coupled
dots is desirable, since in all the above theories, the shape of
the resonant peak is found to be Lorentzian. Experimentally
non-Lorentzian lineshapes are found, but these are attributed to
inelastic scattering. It is therefore of interest to ask under
what circumstances non-Lorentzian peaks are found in the absence
of inelastic scattering.

In this paper we consider a DQD system with coherent interdot coupling as well as
with coherent couplings between the dots and leads. We assume that the charging
energies of the
dots are negligible or can be treated as constant shifts\cite{ref:vaart} and therefore
the system can be described by an effective single particle model.
(For studies of the electronic correlations in DQDs see 
Refs.~\onlinecite{ref:aguado,ref:aono,ref:lopez,ref:orellana,ref:bulka}.)
We note that experimental results  on DQD molecules are often explained
with such simple effective single
particle models\cite{ref:pioro,ref:hatano,ref:graeber,ref:huttel}, which
usually assume that couplings of the dots  to the leads can be neglected.
We  first show  how the finite interdot and dot-lead couplings  affect
the lineshape of the zero bias conductance resonance.
Our results are non-perturbative and  take into account all orders of
interdot and dot-lead tunnel processes.
We also consider   the finite temperature
conductance and present  results for the  temperature dependence of the peak height of the
conductance resonance as well as its broadening.
We show that in our model the non-thermal broadening effect in resonant transport
can be observed even in the zero-bias limit, if the dot-lead coupling strength exceeds the
interdot coupling $t_c$.

The paper is organised  as follows. In Sec. II we introduce a very
general description of coupled dots and derive the zero
temperature transmission formula. In Sec. III we discuss the
properties of the zero-temperature and zero-bias conductance as
functions  of the energy levels of the two dots and of the various
couplings in the systems. In Sec. IV we derive a finite
temperature conductance formula  for the DQD system and discuss
temperature-dependent transport properties of the system.

\section{Model}

\begin{figure}[ht]
\includegraphics[scale=0.35]{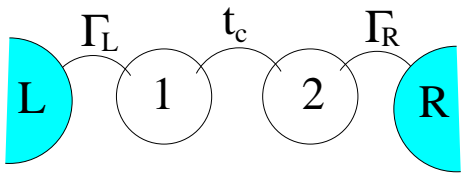}
\caption{Double quantum dot coupled to left (L) and right (R)
leads with interdot coupling $t_c$.
\label{fig:system}}
\end{figure}

We consider two quantum dots coupled to left and right leads with tunnel coupling
$\Gamma_L$ and  $\Gamma_R$, respectively as shown in Fig.~\ref{fig:system}(a).
 The interdot tunnelling coupling is denoted
by $t_c$ and it is assumed that only one energy level
in each dot is relevant.
The intradot as well as the interdot Coulomb interactions are neglected.

As shown in appendix \ref{trans-form-deriv}, the electron transmission coefficient
$T_{dd}(E)$ of the double dot system can be written
\begin{equation}
T_{dd}(E)=\frac{4 \Gamma_L \Gamma_R t_c^2}{|E-E_{+}|^2\,\,|E-E_{-}|^2}.
\label{eq:transmission}
\end{equation}
Here  $E_{\pm}$  are the poles of the transmission:
\begin{equation}
E_{\pm}=\overline{\varepsilon}+ i\,\overline{\Gamma}\pm
\sqrt{(\Delta\varepsilon + i \Delta \Gamma)^2+t_c^2},
\label{eq:poles}
\end{equation}
where
$\overline{\varepsilon}=(\tilde\varepsilon_1+\tilde\varepsilon_2)/2$,
$\overline{\Gamma}=(\Gamma_L+\Gamma_R)/2$ is the total coupling
strength to the leads, while
$\Delta\varepsilon=(\tilde\varepsilon_1-\tilde\varepsilon_2)/2$,
$\Delta{\Gamma}=(\Gamma_L-\Gamma_R)/2$ are the asymmetries of the
dot energies and the couplings to the leads, respectively.

Having obtained the transmission function $T_{dd}(E)$,   the linear conductance
at finite temperature is given by
\begin{equation}
\mathcal{G}_{T}=\frac{2e^2}{h}\int_{0}^{\infty}T_{dd}(E)
\left(-\frac{\partial f_0(E)}{\partial E}\right)\mathrm{d}E,
\label{eq:finite_T_trans}
\end{equation}
where $f_0(E)=[1+\exp((E-\mu)/k_B\,T)]^{-1}$ is the equilibrium Fermi distribution,
$\mu$ being the chemical potential of the leads.
For $k_B\,T\ll E_F$ we can take $\mu \approx E_F$ and extend the lower bound of
the integration to $-\infty$ in Eq.~(\ref{eq:finite_T_trans}). The resulting integral
then can be calculated by contour integration.
 The zero temperature transmission $T_{dd}(E)$ has only simple poles if
 $|\Delta\Gamma|\neq t_c$, $\Delta\varepsilon\neq 0$
and the finite temperature conductance reads:
\begin{widetext}
\begin{equation}
\mathcal{G}_{T}=\frac{2e^2}{h}\frac{\Gamma_L \Gamma_R\, t_c^2}{\pi (k_B\,T)^4}
\left(\frac{1}{(\omega_{+} - \omega_{-})}\left[\frac{1}{\Im\omega_+\,(\omega_{+}-\omega_{-}^*)}
 \psi^{(1)}\left(\frac{1}{2}-\frac{i \omega_{+}}{2\pi}\right)+
 \frac{1}{\Im\omega_-\,(\omega_{+}^*-\omega_{-})}
 \psi^{(1)}\left(\frac{1}{2}-\frac{i \omega_{-}}{2\pi}\right)
 \right]+c.c.\right)
 \label{eq:polygammas}
\end{equation}
\end{widetext}
where $\psi^{(1)}(z)$ is the first polygamma
function\cite{ref:abramowitz}, 
 $\omega_{\pm}=(E_{\pm}-E_F)/k_B T$
(here $E_{\pm}$ are the poles of $T_{dd}(E)$ in the upper half complex plain,
given by Eq.~(\ref{eq:poles})),
$\Im$ denotes the imaginary part, and $^{*}$ stands
for complex conjugation. In case of $\Delta\varepsilon=0$ and
$|\Delta\Gamma|=t_c$, the transmission  $T_{dd}(E)$ has second order
poles and therefore $\mathcal{G}_{T}$ is given by
\begin{widetext}
\begin{equation}
\mathcal{G}_{T}=\frac{e^2}{h}\frac{\Gamma_L \Gamma_R\, t_c^2}{\pi k_B\,T}
\frac{1}{(\Im\omega)^3}
\left[
\psi^{(1)}\left(\frac{1}{2}-\frac{i \omega}{2\pi}\right)
-\frac{\Im\omega}{2\pi}
\psi^{(2)}\left(\frac{1}{2}-\frac{i \omega}{2\pi}\right)+c.c.
\right].
\label{eq:polygammas-secpoles}
\end{equation}
\end{widetext}
Here $\omega=(\overline{\varepsilon}-E_F+i\overline{\Gamma})/k_B T$
and  $\psi^{(2)}(z)$ is the second polygamma function\cite{ref:abramowitz}.

In what follows we briefly discuss the properties of the zero temperature
conductance with emphasis on the effect of couplings $\Gamma_L, \Gamma_R$
and $t_c$. The understanding of the zero temperature case then
helps us  to  interpret the finite temperature behaviour of
the conductance in Sec.~IV .

\section{Zero temperature linear conductance}
From Eq.~\ref{eq:finite_T_trans}, the linear conductance
$\mathcal{G}_0$ at zero temperature is given by the Landauer
formulae\cite{ref:datta-book}
\begin{equation}
\mathcal{G}_0=\frac{2 e^2}{h} T_{dd}(E_F).
\label{eq:zero_T_cond}
\end{equation}
Using   Eq.~(\ref{eq:transmission}) and (\ref{eq:poles})  one can
obtain an explicit expression for $\mathcal{G}_0$ as a function of
$E_F$ and the characteristic energies  of the DQD system:
$\tilde\varepsilon_1$, $\tilde\varepsilon_2$, $\overline{\Gamma}$,
$\Delta\Gamma$ and $t_c$. For $\Delta\Gamma=0$ this  expression
agrees with the result  obtained in Ref.~\onlinecite{ref:guevara}
for serially connected dots, but otherwise also describes the case
when $\Delta\Gamma\neq0$. In an experiment  $E_F$ would be kept
fixed and  the energy levels of the dots as well as the
tunnellings   to the leads  would be changed by side and top
gates. Therefore in what follows  without  loss of generality we
can set $E_F=0$.

\begin{figure}[hbt]
\vspace*{-5mm}
\includegraphics[scale=0.3, angle=-90]{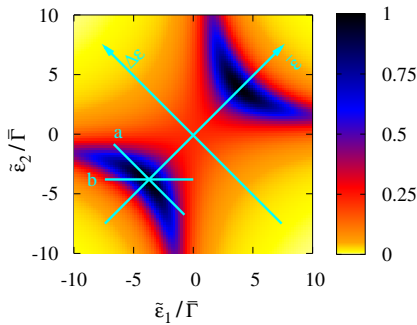}
\vspace*{-3mm} \caption{Zero temperature conductance at $E=E_{F}$ as a function
of $\tilde\varepsilon_1$, $\tilde\varepsilon_2$  for $t_c \gg
\Lambda$. For details see text. \label{fig:zeroTcond3d1}}
\end{figure}

\begin{figure}[hbt]
\vspace*{-3mm}
\includegraphics[scale=0.3, angle=-90]{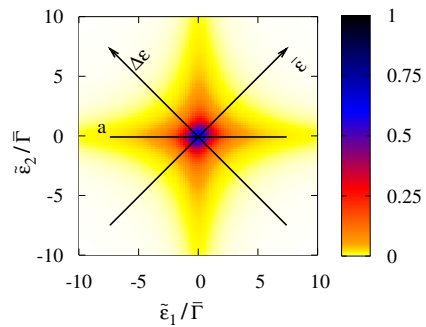}
\vspace*{-2mm} \caption{Zero temperature conductance at $E=E_{F}$ as a function
of $\tilde\varepsilon_1$, $\tilde\varepsilon_2$  for $t_c \ll
\Lambda$. For details see text. \label{fig:zeroTcond3d2}}
\end{figure}

Our aim is to understand the properties of the conductance in the
($\tilde\varepsilon_1$, $\tilde\varepsilon_2$) plane [or
equivalently in ($\overline{\varepsilon}$, $\Delta\varepsilon$)
plane, where
 $\overline{\varepsilon}$ and  $\Delta\varepsilon$ are defined
 after Eq.~(\ref{eq:poles})].
It follows from Eq.~(\ref{eq:transmission}) that depending on the
ratio $t_c^2/\Lambda$, where
$\Lambda=\sqrt{(\Gamma_L^2+\Gamma_R^2)/2}$, the conductance has
either one or two maxima in the ($\tilde\varepsilon_1$,
$\tilde\varepsilon_2$) plane. As shown in
Fig.~\ref{fig:zeroTcond3d1}, for  $t_c \gg \Lambda$ the
conductance is enhanced in two boomerang shaped regions, while for
$t_c \ll \Lambda$ there is one maximum  in the conductance at
$\tilde\varepsilon_1=\tilde\varepsilon_2=E_F$ [see
Fig.~\ref{fig:zeroTcond3d2}].

To see the dependence of the conductance on the energy levels of
the dots and on the various couplings in the system, we now
consider certain directions in the ($\tilde\varepsilon_1$,
$\tilde\varepsilon_2$) plane and study the cross sections of the
conductance along these directions. Let us first assume that the
levels $\tilde\varepsilon_1$, $\tilde\varepsilon_2$ of the two
dots are kept aligned, i.e. $\Delta\varepsilon=0$ and we study
$\mathcal{G}_0$ along the $\overline{\varepsilon}$  axis.

\begin{figure}[ht]
\includegraphics[scale=0.4]{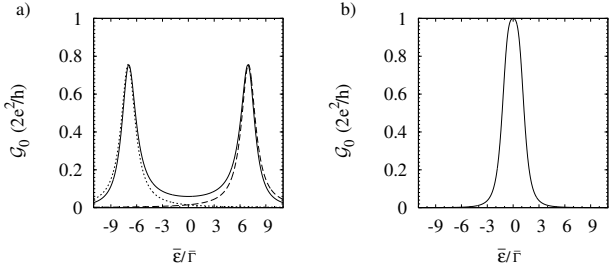}
\caption{ (a) Conductance for $t_c \gg \Lambda$ along the
$\overline{\varepsilon}$ axis using Eq.~(\ref{eq:transmission})
(solid line). The approximation given by
Eq.~(\ref{eq:approxT0Lorentz}) is shown with dashed lines.
We used $t_c/\overline{\Gamma}=7$ and $\Delta\Gamma/\overline{\Gamma}=0.5$.
(b) The transmission for
$t_c =\sqrt{\Gamma_L\Gamma_R} < \Lambda$ given by Eq.~(\ref{eq:transT0oneres}).
The parameters are
$t_c/\overline{\Gamma}=0.954$ and $\Delta\Gamma/\overline{\Gamma}=0.3$.
For details see text.
\label{fig:zeroTcond2}}
\end{figure}

If $t_c > \Lambda$, upon varying
$\overline{\varepsilon}$ two resonances occur in the conductance
 [see  Figs.~\ref{fig:zeroTcond3d1},~\ref{fig:zeroTcond2}(a)] at energies
\begin{equation}
\overline{\varepsilon}_{\pm}=\pm\sqrt{t_c^2-\Lambda^2},
\label{eq:bond-antibond}
\end{equation}
corresponding to symmetric (-) and antisymmetric (+) molecular states.
Note that in an isolated double dot molecule where $\Gamma_L=\Gamma_R=0$,
 the energies of the symmetric (antisymmetric)  states
 are $\overline{\varepsilon}_{\pm}^{\,0}=\pm t_c$.
Thus Eq.~(\ref{eq:bond-antibond}) shows that  connection to the
leads produces  level attraction.
Since $\Lambda=\sqrt{\overline{\Gamma}^2+\Delta\Gamma^2}$,
the magnitude of the level attraction
depends both on the total coupling strength $\overline{\Gamma}$ and on the
asymmetry  $\Delta\Gamma$ of the couplings.
Similar level attraction has been reported in
Ref.~\onlinecite{ref:guevara} for QDs attached in parallel to the leads and
in Ref.~\onlinecite{ref:kubala1} for an Aharonov-Bohm ring with a quantum dot in each of
its arms. It can also be  shown that $\mathcal{G}_0$  reaches the quantum limit
$2e^2/h$ at energies given by Eq.~(\ref{eq:bond-antibond})
only if $\Delta\Gamma=0$, i.e. for symmetric couplings to the leads.
For $t_c\gg\Lambda$, when the two resonances are well separated,
the lineshape around
$\overline{\varepsilon}=\pm\sqrt{t_c^2-\Lambda^2}$ is
approximately a Lorentzian of linewidth $\overline{\Gamma}$:
\begin{equation}
\mathcal{G}_0(\overline{\varepsilon})\approx \frac{2 e^2}{h}
\frac{t_c^2}{(t_c^2-(\overline{\Gamma}^2/4+\Delta\Gamma^2))}
\frac{\Gamma_L\Gamma_R}
{[(\overline{\varepsilon}\pm\sqrt{t_c^2-\Delta\Gamma^2})^2+\overline{\Gamma}^2]}.
\label{eq:approxT0Lorentz}
\end{equation}
Regarding the enhancement of the conductance in a boomerang-shape
areas in the $(\tilde\varepsilon_1, \tilde\varepsilon_2)$ plane
(see Fig.~\ref{fig:zeroTcond3d1}), it is easy to prove that if
$\Delta\Gamma=0$, for a given value of  $\Delta\varepsilon$, the
transmission $T_{dd}$ has maximum at $\overline{\varepsilon}=\pm
\sqrt{\Delta\varepsilon^2+t_c^2-\Gamma^2}$. This is the equation
of a hyperbola in the $(\overline{\varepsilon},\Delta\varepsilon)$
plane and also helps  to understand the observed structure of the
conductance when $t_c \gg |\Delta\Gamma|\neq 0$, which is the case
in Fig.~\ref{fig:zeroTcond3d1}.

The separation between the two resonances decreases as  $\Lambda$
is increased while keeping $t_c$ fixed. Finally, the two
resonances merge when  $\Lambda=t_c$ meaning that due to the
coupling to the leads, the energies of the symmetric and
antisymmetric states become degenerate at this value of $\Lambda$.
For $\Lambda > t_c$ the conductance then has only one resonance at
energy $\overline{\varepsilon}=E_F$ [see
Figs.~\ref{fig:zeroTcond3d2}, ~\ref{fig:zeroTcond2}(b)], and from
Eqs.~(\ref{eq:transmission}) and (\ref{eq:poles}), has the form
\begin{equation}
\mathcal{G}_0(\overline{\varepsilon})=\frac{2 e^2}{h}
\frac{4\Gamma_L \Gamma_R\, t_c^2}{[(\overline{\varepsilon}+\tilde{t}_c)^2+\Gamma_{+}^2]
[(\overline{\varepsilon}-\tilde{t}_c)^2+\Gamma_{-}^2]}.
\label{eq:transT0oneres}
\end{equation}
Here $\tilde{t}_c=\sqrt{t_c^2-\Delta\Gamma^2}$, $\Gamma_{\pm}=\overline{\Gamma}$
if  $|\Delta\Gamma| \le t_c \le \overline{\Gamma}$  and one can show that
the conductance on resonance
$\mathcal{G}_0(\overline{\varepsilon}=E_F)$ is smaller than $2e^2/h$ except for
the special case $t_c=\sqrt{\Gamma_L \Gamma_R}$\,\cite{ref:matveev2} [see Fig.~\ref{fig:zeroTcond2}(b)].
On the other hand, if
$t_c < |\Delta\Gamma|<\overline{\Gamma}$, we  have
 $\tilde{t}_c=0$ and $\Gamma_{\pm}=\overline{\Gamma}\pm \sqrt{\Delta\Gamma^2-t_c^2}$
in Eq.~(\ref{eq:transT0oneres}), and we find that
$\mathcal{G}_0(\overline{\varepsilon})<2e^2/h$
for all $\overline{\varepsilon}$.
Thus for $t_c<\Lambda$
the lineshape of the resonance can again be approximated by  a Lorentzian around
 $\overline{\varepsilon}=E_F$ but  for larger $|\overline{\varepsilon}|$
  it decreases as $\sim 1/ \overline{\varepsilon}^4$.

Further   understanding of the properties of the conductance can
be gained by considering  $\Delta\varepsilon\neq 0$, i.e. finite
detuning between the levels of the (isolated) dots. In principle
one could follow any path in the $(\tilde\varepsilon_1,
\tilde\varepsilon_2)$ plane to study the effect of finite $
\Delta\varepsilon$, but let us consider two simple yet important
cases. Let us first assume that $\overline{\varepsilon}$ is kept
fixed at the value where the conductance is largest and we  vary
only  $\Delta\varepsilon$, ie we study the conductance parallel to
the $\Delta\varepsilon$ axis. Analytical progress can be made most
easily for $\Delta\Gamma=0$, i.e. $\Gamma_L=\Gamma_R=\Gamma$.

When $t_c > \Lambda=\Gamma$ and
$\overline{\varepsilon}=\overline{\varepsilon}_{\pm}$
i.e. $\overline{\varepsilon}$ equals the
energy of the symmetric (antisymmetric) state, the conductance as a
function of  $\Delta\varepsilon$ reads
\begin{equation}
\mathcal{G}_0(\Delta\varepsilon)=\frac{2 e^2}{h}
\frac{4\Gamma^2 t_c^2}{\Delta\varepsilon^2 (\Delta\varepsilon^2+4\Gamma^2)+4\Gamma^2 t_c^2}.
\label{eq:cond_T0_1}
\end{equation}
(This corresponds to taking the cross section of $T_{dd}(E)$ along
the line denoted by "a" in Fig.~\ref{fig:zeroTcond3d1}.) We see
that the lineshape  is basically a Lorentzian for
$\Delta\varepsilon\ll\Gamma$, but falls more rapidly for
$\Delta\varepsilon  \gtrsim\Gamma$. The  full width at half
maximum (FWHM) is ${\Delta\varepsilon_W=2\sqrt{2
\Gamma(\sqrt{\Gamma^2+t_c^2}-\Gamma)}}$,
 which simplifies to
$\Delta\varepsilon_W\approx 2 \sqrt{2\Gamma(t_c-\Gamma)}$ for $t_c\gg \Gamma$.

Another obvious way  to study the effect of finite
$\Delta\varepsilon$ is to fix one of the energies, (e.g.
$\tilde\varepsilon_2$) and vary only $\tilde\varepsilon_1$.
 In contrast with the previous example, this means
 that both  $\overline{\varepsilon}$ and $\Delta\varepsilon$ are  being varied.
For $\Delta\Gamma=0$, keeping 
$\tilde\varepsilon_2=\overline{\varepsilon}_{\pm}$
 fixed and  varying only $\tilde\varepsilon_1$ (see line "b" in Fig.~\ref{fig:zeroTcond3d1})
 yields a particularly simple result for
 the conductance:
\begin{equation}
\mathcal{G}_0(\Delta\varepsilon)=\frac{2 e^2}{h}\frac{\Gamma^2}{\Delta\varepsilon^2+\Gamma^2}.
\end{equation}
We see that the lineshape in this case is a simple Lorentzian which is,
interestingly, independent of $t_c$. The FWHM reads $\Delta\varepsilon_{W}=2\Gamma$.

On the other hand, if  $t_c < \Lambda=\Gamma$, one can easily show that for
 fixed $\overline{\varepsilon}=E_F$  the conductance along the
 $\Delta\varepsilon$ axis reads
 \begin{equation}
\mathcal{G}_0(\Delta\varepsilon)=\frac{2 e^2}{h}
\frac{4\Gamma^2 t_c^2}{(\Delta\varepsilon^2+\Gamma^2+t_c^2)^2},
\label{eq:cond_T0_2}
\end{equation}
meaning that  the lineshape is approximatelly  a Lorentzian for
$\Delta\varepsilon\ll\sqrt{\Gamma^2+t_c^2}$ and  the
  FWHM is  $\Delta\varepsilon_W=2\sqrt{(\sqrt{2}-1)(\Gamma^2+t_c^2)}$.
Finally, if  we  keep  $\tilde\varepsilon_2$  aligned with $E_F$
and vary only
 $\tilde\varepsilon_1$ (see  line "a" in Fig.~\ref{fig:zeroTcond3d2})
 the conductance as a function of $\Delta\varepsilon$ reads
 \begin{equation}
\mathcal{G}_0(\Delta\varepsilon)=\frac{2 e^2}{h}
\frac{t_c^2}{\Delta\varepsilon^2+\frac{\Gamma^2}{4}\left(1+\frac{t_c^2}{\Gamma^2} \right)^2}.
\label{eq:cond_T0_3}
\end{equation}
Thus the lineshape is a Lorentzian  which means  a weaker
$\Delta\varepsilon$  dependence of the conductance than in
Eq.~(\ref{eq:cond_T0_2}). This is readily seen in
Fig.~\ref{fig:zeroTcond3d2} as high conductance ridges extending
further along the  $\tilde\varepsilon_1$ and $\tilde\varepsilon_2$
direction than along the $\Delta\varepsilon$ axis.
 The width of the resonance is $\Delta\varepsilon_{W}=(\Gamma^2+t_c^2)/\Gamma$.

\begin{figure}[ht]
\includegraphics[scale=0.6]{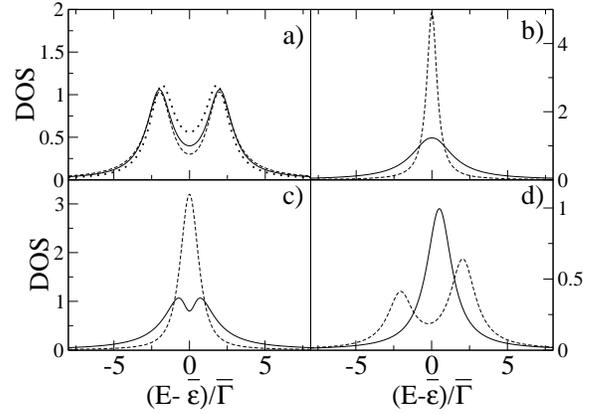}
\caption{ (a) Local DOS $\rho_1(E)$  for $t_c/\overline{\Gamma}=2$, 
$\Delta\Gamma=0$  (solid line), and  $\rho_1(E)$, $\rho_2(E)$ for  $\Delta\Gamma/\overline{\Gamma}=0.6$ (dashed and dotted lines, respectively).
 (b) Local DOS $\rho_1(E)$, $\rho_2(E)$ (solid and dashed lines, respectively) 
for $\Delta\Gamma/\overline{\Gamma}=0.6$,
$t_c/\overline{\Gamma}=0.1$, and (c) for $t_c/\overline{\Gamma}=0.5$.
(d) $\rho_1(E)$ for $\Delta \varepsilon/\overline{\Gamma}=0.5$,
$t_c/\overline{\Gamma}=0.1$ (solid line) and $t_c/\overline{\Gamma}=2.0$
 (dashed line). 
\label{fig:localdos}}
\end{figure}

To complete our analysis of the zero temperature physics of our  model, 
we now briefly discuss the 
local density of states (LDOS) in each of the dots. The LDOS of dot 1 (2) 
can be obtained  from the diagonal matrix elements of the Green's function $G_{DD}$ 
of the DQD system (see Eq.~\ref{GBB}):
\begin{equation}
\rho_i(E)=-\frac{1}{\pi}{\rm Im}(G_{DD})_{ii}, \hspace*{0.5cm} i=1,2.
\end{equation}   
Our main findings are summarized in Fig.\ref{fig:localdos}.
If the levels of the two dots are equal 
$\tilde{\varepsilon}_1=\tilde{\varepsilon}_2=\overline{\varepsilon}$  
then depending on the ratio of the coupling $t_c$ and the asymmetry of the 
dot-lead couplings  $\Delta\Gamma$, one can discern three cases. For 
 $\Delta\Gamma=0$ the LDOS, which is  the same in both dots, is a superposition 
of  two Lorentzians centered on  energies $E=\overline{\varepsilon}\pm t_c$ 
(see Fig.~\ref{fig:localdos}(a)). Upon decreasing the  
ratio $t_c/\Delta\Gamma $ the twin peak structure of the LDOS remains as long
as $t_c/\Delta\Gamma>1$ with the LDOS of  the dot  with smaller dot-lead coupling 
being is larger than that of the other dot's. If the 
coupling between the dots is weak so that $t_c/\Delta\Gamma\ll 1$ 
[as in Fig.~\ref{fig:localdos}(b)] then the LDOS in each of the dots 
is basically a Lorentzian of  width approximately 
that of the corresponding 
couplings $\Gamma_L (\Gamma_R)$.
In the intermediate regime of $ t_c/\Delta\Gamma \lesssim 1$ an interesting 
difference in the LDOS of the two dots can be observed 
(Fig.~\ref{fig:localdos}(c)).  The dot having stronger 
coupling to the corresponding lead has a double peaked LDOS while 
the other's LDOS is  single peaked.  If the levels of the dots are detuned 
($\Delta\varepsilon\neq 0$)  we consider the case when $\Delta\Gamma=0$.
The LDOS then depends on the ratio  $t_c/\Delta\varepsilon$. If the detuning
of the levels is large compared  to the coupling 
i.e. when $t_c/\Delta\varepsilon\ll 1$
the LDOS in  dot 1 (2) is peaked at $E=\tilde{\varepsilon}_1$ 
($\tilde{\varepsilon}_2$) and
has little overlap with the LDOS of the other dot. In the opposite limit of 
$t_c/\Delta\varepsilon\gg 1$ the LDOS is a superposition of two 
Lorentzians  peaked at the energies of the 
molecular states $E=\overline{\varepsilon}\pm t_c$.

\section{Finite temperature linear conductance}

\begin{figure}[htb]
\includegraphics[scale=0.4]{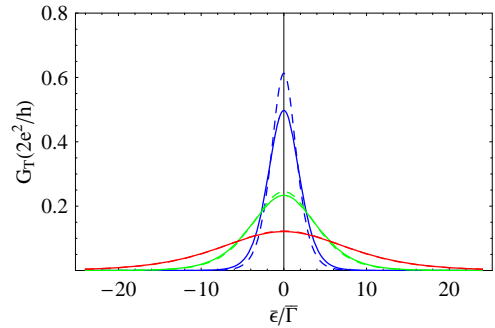}
\caption{Conductance (in units of $2e^2/h$) as a function of  $\overline{\varepsilon}$
for three different temperatures: $k_B T/\overline{\Gamma}=1,2.5,5$
 (blue, green and red lines, respectively).
Curves with solid lines show the exact result of Eq.~(\ref{eq:polygammas}),
the curves with dashed lines are calculated using  Eq.~(\ref{eq:lineshapeT-oneres}).
\label{fig:peakT-oneres}}
\end{figure}

We now consider the properties of the conductance for finite
temperature. Although the formulas given in
Eqs.~(\ref{eq:polygammas}) and (\ref{eq:polygammas-secpoles}) are
not easily readable, insight into their physical content can again
be gained by scanning along certain direction in the
$(\tilde\varepsilon_1, \tilde\varepsilon_2)$ plane, as  in the
previous section. Let us focus on the
 case when $|\Delta\Gamma|\neq t_c$, $\Delta\varepsilon\neq 0$  so that the
 finite temperature conductance is given by Eq.~(\ref{eq:polygammas}) and
 let us first address
 the question of the temperature dependence of conductance in case of $t_c < \Lambda$,
$\Delta\varepsilon=0$. As before, we assume that $E_F$ is kept constant and
$\overline{\varepsilon}$ is varied. The energy and temperature dependence
of $\mathcal{G}_{T}$  as given by Eq.~(\ref{eq:polygammas}) is shown
by  solid lines in Fig.~\ref{fig:peakT-oneres}.
As the temperature increases, the resonance gradually broadens and its height
 decreases monotonically with $T$. This behaviour can be
 understood in the limit of 
$k_B T \gg  \Lambda > t_c$ (with $k_B T\ll E_F$). In this case
around $\overline{\varepsilon}=E_F$, where the conductance is
significant, one finds that
$|\omega_{\pm}|/{2\pi}=|E_{\pm}-E_F|/2\pi k_B T \ll 1$ and
therefore the expansion of the polygamma functions around $1/2$
can be used:
\begin{equation}
\psi^{(1)}(\frac{1}{2}+\frac{i\,z}{2\pi})\approx \frac{\pi^2}{2}+
\frac{i \psi^{(2)}(1/2)z}{2\pi}-\frac{\pi^4 z^2}{8}+\mathcal{O}(z^3),
\,\, |\textnormal{arg}z|<\pi.
\label{eq:polygamma_expansion}
\end{equation}
Substituting this expansion into Eq.~(\ref{eq:polygammas}), we find that
to leading
order in  $T$ the  peak height $\mathcal{G}_{T,\rm{max}}$
decreases monotonically with the temperature:
\begin{multline}
\mathcal{G}_{T,\rm{max}}\approx \frac{2 e^2}{h} \frac{\pi}{k_B T}
\frac{\Gamma_L\Gamma_R}{\Gamma_L+\Gamma_R}\frac{t_c^2}{(\Gamma_L\Gamma_R+t_c^2)}\\
 \mbox{if  }  k_B T \gg \Lambda > t_c \mbox{  and  } \Delta\varepsilon = 0.
\label{eq:GpeakT-oneres}
\end{multline}
A similar $\sim 1/ T$  decay can be found for the conductance of a
single dot\cite{ref:beenakker}.
However, the double dot result differs from the result for
a single dot by a factor of $2 t_c^2/(\Gamma_L\Gamma_R+t_c^2)$, i.e.
the degeneracy of the symmetric and antisymmetric levels does not
simply contribute a factor of $2$ to the conductance
as one could na{\"\i}vly expect, but also a factor
of $t_c^2/(\Gamma_L\Gamma_R+t_c^2)$.
As shown in Fig.~\ref{fig:peakT-oneres}, for $k_B T \gg \Lambda >
t_c$ the conductance is approximated by
\begin{multline}
\mathcal{G}_{T}(\overline{\varepsilon})\approx\mathcal{G}_{T,\rm{max}}
\cosh^{-2}\left(\frac{\overline{\varepsilon}}{2 k_B T}\right)\\
\mbox{if  }  k_B T \gg \Lambda > t_c \mbox{  and  } \Delta\varepsilon = 0
\label{eq:lineshapeT-oneres}
\end{multline}
so that apart from the amplitude $\mathcal{G}_{T,\rm{max}}$ the
lineshape is the same as for a single dot. It follows from
Eq.~(\ref{eq:lineshapeT-oneres}) that the FWHM is a linear
function of the temperature with a slope of
$2\,\textrm{acosh}(\sqrt{2})$\cite{ref:kotlyar,ref:beenakker}. We
see that for high temperatures  along the $\overline{\varepsilon}$
axes, the conductance of a double dot is similar to that of a
single dot with degenerate energy levels.

\begin{figure}[htb]
\includegraphics[scale=0.4]{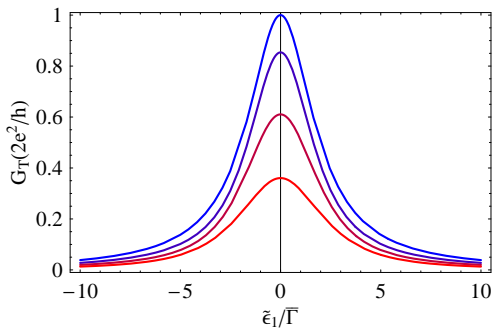}
\caption{Conductance (in units of $2e^2/h$) as a function of
$\tilde\varepsilon_1$ 
for  $\tilde\varepsilon_2=E_F$ fixed using four different
temperatures: $k_B T/\overline{\Gamma}=0,0.5,1.0,2.0$ (light blue,
dark blue, brown and red lines, respectively).
\label{fig:nonTbroad-oneres}}
\end{figure}

\begin{figure}[hbt]
\includegraphics[scale=0.35]{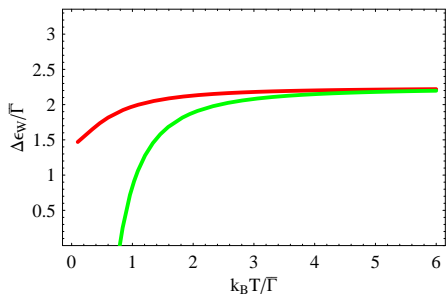}
\caption{FWHM  as a function of temperature,  along the $\Delta\varepsilon$ axis.
The red curve shows the numerically obtained FWHM using Eq.~(\ref{eq:polygammas}),
the green curve is the function given by Eq.~(\ref{eq:fwhmT-oneres-1}).
We used  $t_c/\Gamma=0.5$.
\label{fig:fwhm-oneres}}
\end{figure}

In contrast to the above behaviour, if we keep
$\overline{\varepsilon}=E_F$ fixed or  $\tilde\varepsilon_2=E_F$
fixed and calculate the transmission as a function of
$\Delta\varepsilon$ for different temperatures, we see that the
resonance peak is not broadened by temperature
(Fig.~\ref{fig:nonTbroad-oneres}).  Indeed, if we numerically
calculate the FWHM as a function of $T$ using
Eq.~(\ref{eq:polygammas}), we see in Fig.~\ref{fig:fwhm-oneres}
(red curve) that in both cases after an initial increase of FWHM
in the regime of  $k_B T \lesssim \overline{\Gamma}$, the width of
the resonance approaches a constant value as the temperature  is
further  increased.
For $\Delta\Gamma=0$ using the  expansion shown in Eq.~(\ref{eq:polygamma_expansion})
we find that for fixed $\overline{\varepsilon}=E_F$ the FWHM is given by
\begin{equation}
\Delta\varepsilon_{W,T}\approx 2\sqrt{\Gamma^2+t_c^2}-\frac{(\Gamma^2+t_c^2)^{3/2}}{2 (k_B T)^2},
\label{eq:fwhmT-oneres-1}
\end{equation}
therefore  the high temperature value of the FWHM only  weakly  increases
with  temperature  since $(\Gamma^2+t_c^2)^{3/2}/(2 k_B T)^2\ll 1$
 [see Fig.~\ref{fig:fwhm-oneres}].
The ratio of the zero and finite temperature FWHMs is
$\Delta\varepsilon_{W,T}/\Delta\varepsilon_{W}\approx 1.55$ [see
$\Delta\varepsilon_{W}$ after Eq.~(\ref{eq:cond_T0_2})]. On the
other hand, if  e.g. $\tilde\varepsilon_2=E_F$ is fixed and
 $\tilde\varepsilon_1$ is changed, a similar calculation yields
\begin{equation}
\Delta\varepsilon_{W,T}\approx 2\sqrt{\Gamma^2+t_c^2}-\frac{ (\Gamma^2+t_c^2)^{3/2}}{(k_B T)^2},
\label{eq:fwhmT-oneres-2}
\end{equation}
thus $\Delta\varepsilon_{W,T}$ again shows a weak temperature dependence for
$k_B T \gtrsim \Gamma$
and compared to the $T=0$ case  we see that
$\Delta\varepsilon_{W,T}/\Delta\varepsilon_{W} \approx 2 \Gamma / \sqrt{\Gamma^2+t_c^2}$,
which for  $\Gamma \gg t_c$ gives
$\Delta\varepsilon_{W,T}/\Delta\varepsilon_{W}\approx 2$.
We find therefore that for both scenarios the FWHM increases as a function
of temperature if $k_B T \lesssim \Gamma$ but for
 $k_B T \gg \Gamma$ it approaches a constant value of
$2\sqrt{\Gamma^2+t_c^2}$. This is a remarkable  non-thermal
broadening effect and is a central result of our paper: while the
peak height decreases monotonically with temperature [see
Eq.~(\ref{eq:GpeakT-oneres})], the width of the peak, when
changing only $\Delta\varepsilon$  or $\tilde\varepsilon_1$,
approaches a constant value, as shown in
Eqs.~(\ref{eq:fwhmT-oneres-1}), (\ref{eq:fwhmT-oneres-2}) and in
Fig.~\ref{fig:fwhm-oneres}. We emphasize that although for 
the analytic calculation we assumed $\Delta\Gamma=0$, our numerical 
results show that the nonthermal broadening is also present
 for a finite difference in the couplings to the leads, i.e. 
 when  $\Delta\Gamma\neq 0$.
One can also show that for large $T$, where the FWHM is approximately
constant, the lineshape of the resonance is approximately a
Lorentzian:
\begin{multline}
\mathcal{G}_{T}(\Delta\varepsilon)=\frac{2e^2}{h}\frac{\pi\Gamma}{k_B T}
\frac{ t_c^2}{\Delta\varepsilon^2+\Gamma^2+t_c^2}\\
\mbox{if  }  k_B T \gg \Gamma > t_c,\mbox{ and }\Delta\Gamma=0.
\end{multline}
This result holds for both of the scenarios discussed i.e. either
$\overline{\varepsilon}$ or $\tilde\varepsilon_2$ being fixed
while  $\Delta\varepsilon$ is varied.

\begin{figure}[htb]
\includegraphics[scale=0.4]{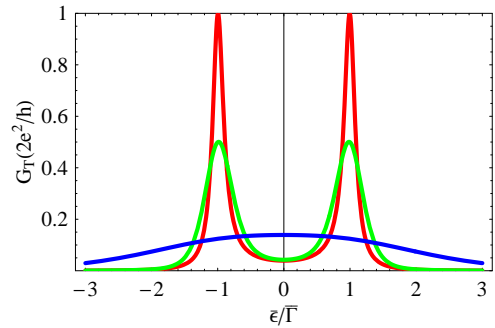}
\caption{Conductance as a function of  $\overline{\varepsilon}$
for three different temperatures: $k_B T/\overline{\Gamma}=0,1,8$.
(red, green and blue lines, respectively). The ratio of $t_c$ and $\overline{\Gamma}$ is
$t_c/\overline{\Gamma}=10$, while ${\Delta\Gamma=0}$.
\label{fig:peakT-twores}}
\end{figure}

Let us now consider the case of $t_c > \Lambda$.
The temperature dependence of $\mathcal{G}_{T}$ as given
by Eq.~\ref{eq:polygammas} along the  $\overline{\varepsilon}$ axis  is shown
in Fig.~\ref{fig:peakT-twores}.
We see that the conductance peaks  gradually broaden
and their height  decreases with increasing temperature
and finally they merge into a single peak for $k_B T \gtrsim t_c$.
The maximum of the  conductance can then be found at $\overline{\varepsilon}=E_F$
and  increasing  the temperature further to the regime of
$k_BT \gg t_c > \Lambda$
this peak behaves  the same way as
the case where we  assumed $t_c<\Lambda$.
This  occurs because in the high temperature limit $k_B T$ is the largest
energy scale in the system and  $|\omega_{\pm}| \ll k_B T$. Hence the expansion
shown in Eq.~(\ref{eq:polygamma_expansion}) is
applicable and  leads to the same results as
Eqs.~(\ref{eq:GpeakT-oneres}), (\ref{eq:lineshapeT-oneres}), (\ref{eq:fwhmT-oneres-1})
and (\ref{eq:fwhmT-oneres-2}).

For $t_c \gg k_B T, \Lambda$ however, when one can still observe two distinct peaks in the
conductance (see e.g. the  green curve in Fig.~\ref{fig:peakT-twores})
different approach has to be employed. Let us focus on the
peak at the energy of the symmetric state,
i.e. $\overline{\varepsilon}_{-}=-\sqrt{t_c^2-\Lambda^2}$.
(Analogous considerations can be made around the energy $\overline{\varepsilon}_{+}=\sqrt{t_c^2-\Lambda^2}$
of the antisymmetric state.)
If $t_c \gg k_B T, \Lambda$, then for energies  around $\overline{\varepsilon}_{-}$,
one finds that
$|\omega_{+}|/2\pi \ll 1$, but $|\omega_{-}|/2\pi \gg 1$ in Eq.~(\ref{eq:polygammas}).
This implies that in the expression of $\mathcal{G}_T$
the polygamma functions, whose
argument  is $\omega_{-}$ can  be neglected compared with
the other  two terms, which are functions of  $\omega_{+}$. Indeed, from
 the expansion of polygamma functions for large arguments
\begin{equation}
\psi^{(1)}(z)\approx \frac{1}{z}+\frac{1}{2 z^2}+\frac{1}{6 z^3},\,\,\,\,
z\rightarrow\infty,\,\,\, |\textrm{arg} z|<\pi,
\label{eq:polygamma_lowTexp}
\end{equation}
it is clear that the  contribution of  $\psi^{(1)}(1/2-i\omega_{-}/ 2\pi)$ and its
complex conjugate's in Eq.~(\ref{eq:polygammas}) is very small if
$\overline{\varepsilon}\approx \overline{\varepsilon}_{-}$
which  means that for a zero bias measurement
one can consider the symmetric state as a single  resonant level.
The finite temperature conductance
around $\overline{\varepsilon}_{-}$  can be then approximated by
\begin{widetext}
\begin{equation}
\begin{split}
\mathcal{G}_T & \approx \frac{2e^2}{h}\frac{t_c^2}{\pi(k_B T)^2}
\frac{\Gamma_L \Gamma_R}{\Gamma_L+\Gamma_R}\frac{1}{\sqrt{t_c^2-\Delta\Gamma^2}}
\left( \frac{1}{(\omega_{+}-\omega_{-}^*)}
\psi^{(1)}\left(\frac{1}{2}-\frac{i \omega_{+}}{2\pi}\right)+c.c. \right)\\
\mbox{if  } t_c\gg k_B T,\Lambda \mbox{  and  } \Delta\varepsilon=0.
\label{eq:polygammas_symstate}
\end{split}
\end{equation}
\end{widetext}

\begin{figure}[htb]
\includegraphics[scale=0.4]{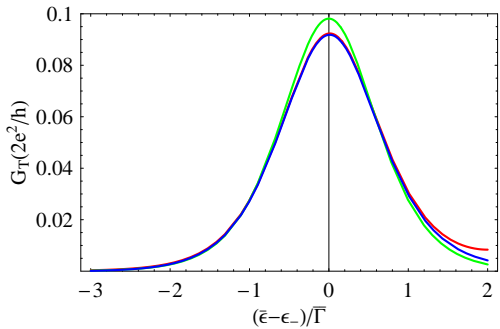}
\caption{Conductance as a function of  $\overline{\varepsilon}$ around the energy
 of the symmetric state $\varepsilon_{-}$ if $t_c \gg k_B T \gg \overline{\Gamma}$.
 The exact result of Eq.~(\ref{eq:polygammas}) is shown with red line, the
 approximation of Eq.~(\ref{eq:polygammas_symstate}) with blue and the
 formula given by Eq.~(\ref{eq:lineshapeT-twores}) with green line.
 Parameters: $k_B T/\overline{\Gamma}=8$, $t_c/k_B T = 5$, $\Delta\Gamma_L=0$.
\label{fig:lineshape-symstate}}
\end{figure}

In case
$k_B T$ is much larger than  $\Lambda$, i.e. for
${t_c \gg k_B T \gg \Lambda}$
 one can use the expansion shown
in Eq.~(\ref{eq:polygamma_expansion}) to obtain  further approximations
of Eq.~(\ref{eq:polygammas_symstate}). For the
peak height we find very similar results to the
case of $t_c<\Lambda$. To leading order the peak height
decreases monotonically with temperature:
\begin{multline}
\mathcal{G}_{T,\rm{max}}\approx \frac{2 e^2}{h} \frac{\pi}{2 k_B T}
\frac{\Gamma_L\Gamma_R}{\Gamma_L+\Gamma_R}\frac{t_c^2}{(\Gamma_L\Gamma_R+t_c^2)}\\
\mbox{if  } t_c\gg k_B T \gg \Lambda \mbox{  and  } \Delta\varepsilon=0.
\label{eq:GpeakT-twores-sym}
\end{multline}
Note that there is  a factor of $1/2$ difference compared to Eq.~(\ref{eq:GpeakT-oneres}),
because the symmetric state behaves as a  single resonant level. One also finds
that the lineshape is rather well approximated by
\begin{multline}
\mathcal{G}_{T}/ \mathcal{G}_{T,\rm{max}}\approx \cosh^{-2} \left(
\frac{\overline{\varepsilon}-\varepsilon_{-}}{2 k_B T}\right)\\
\mbox{if  } t_c\gg k_B T \gg \Lambda \mbox{  and  } \Delta\varepsilon=0.
\label{eq:lineshapeT-twores}
\end{multline}
especially for $\overline{\varepsilon}\ll \varepsilon_{-}$. However,
due to the other resonant level at energy $\varepsilon_{+}$, the lineshape is in fact not
symmetric around  $\varepsilon_{-}$ as Eq.~(\ref{eq:lineshapeT-twores}) suggests.
Comparision of the exact result given by Eq.~(\ref{eq:polygammas}) with the approximations
of Eq.~(\ref{eq:polygammas_symstate}) and Eq.~(\ref{eq:lineshapeT-twores})
is shown in Fig.~\ref{fig:lineshape-symstate}.
A small but noticable deviation of the approximation
given by Eq.~(\ref{eq:lineshapeT-twores})
from the exact result Eq.~(\ref{eq:polygammas}) can indeed be observed
for $\overline{\varepsilon} \gtrapprox \varepsilon_{-}$, while Eq.~(\ref{eq:polygammas_symstate})
gives a better approximation over the whole energy range around $\varepsilon_{-}$.

\begin{figure}[hbt]
 \includegraphics[scale=0.35]{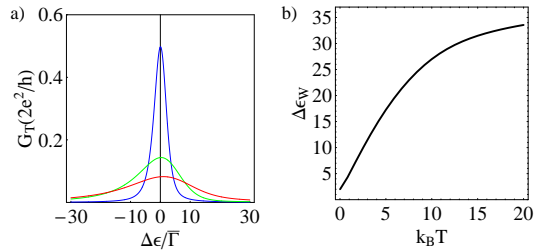}
 \caption{The conductance  as a function of $\Delta\varepsilon$, for three
 different temperatures: $k_B T / \Gamma=1, 5, 10$
(blue, green and red lines, respectively) and fixed
$\tilde\varepsilon_1=\varepsilon_{-}$ (a). The numerically
calculated FWHM using Eq.~(\ref{eq:polygammas}) as a function of
temperature (b). We used  $\Delta\Gamma=0$ and  $t_c/\Gamma =20$.
 \label{fig:fwhm-twores-1}}
 \end{figure}

For the properties of the conductance along the
$\Delta\varepsilon$ axes or along the $\tilde\varepsilon_1$ axes,
 the   characteristic energies $\Delta\varepsilon_{W,T}$ for each case can be
calculated using Eq.~(\ref{eq:polygammas_symstate}) and the expansion
shown in Eq.~(\ref{eq:polygamma_expansion}).
The resulting formulas are however rather complicated
and not too informative.  Numerical calculations shown in
Fig.~\ref{fig:fwhm-twores-1}
clearly indicate that
 the resonance is broadened by temperature in the regime of $t_c \gg k_B T,\Gamma$.

We can conclude therefore that in the studied
system the non-thermal broadening of the conductance resonance occurs
only if $k_B T $ is the largest energy scale.

\section{Conclusions}
In conclusion, we have studied the linear conductance of a double
quantum dot molecule at zero and finite temperatures. We have
found that the coupling of the dots to the leads produces level
attraction,  which depends both an the  total coupling  strength
$\overline{\Gamma}$ and on the asymmetry $\Delta\Gamma$ of the
couplings. We have discussed the properties of the conductance
when the interdot coupling $t_c$ is larger (smaller) than this
level attraction. In particular, at zero temperature we have given
explicit expression for the line shape of the conductance in the
($\tilde\varepsilon_1$, $\tilde\varepsilon_2$) plane along certain
experimentally  important axes. 
Considering the finite temperature conductance we have  discussed
the temperature dependence and the  lineshape of the conductance
along these axes. We have showed that  if the temperature is the
largest energy scale in the system, the conductance resonance,
which arises  due to the detuning of the energy levels of the
quantum dots,  is not broadened by the temperature. Our results
can be relevant for understanding  of those recent experimental
results  where an effective single particle description is
adequate.

\begin{acknowledgments}
Enlightening discussions with Dr.~Andr\'as Csord\'as are gratefully acknowledged.
This work is supported
by E.~C.~Contract No. MRTN-CT-2003-504574 and EPSRC.
\end{acknowledgments}

\appendix
\section{}
\label{trans-form-deriv}

 In this appendix we derive the electron transmission coefficient
 through serially coupled, coherent quantum
dots, connected to multi-channel leads. There are numerous
equivalent approaches to computing transport through such
phase-coherent structures, including recursion methods
 and transfer matrix techniques\cite{ref:colin1,ref:colin2}.
 Here we employ the Green's-function method and notation presented
 in Ref.~\onlinecite{ref:colin3}, in
which the Hilbert space is divided into a sub-space $A$ containing
the external leads and a sub-space $B$ containing the two dots.

We start by considering isolated left and right dots, which are
each described by a single quantum state $\vert f_1>$ and $\vert
f_2>$, with energy levels $\varepsilon_1$ and $\varepsilon_2$
respectively. When these are coupled together by a Hamiltonian
$h_{12}$, the $2\times  2$ Greens function $g_B$ of the coupled dots is
given by
\begin{equation}
g_D^{-1}=\left(%
\begin{array}{cc}
  g_{11}^{-1} & - t_c \\
  - t_c^* & g_{22}^{-1} \\
\end{array}%
\right),\\
\end{equation}
where $ t_c = <f_1\vert h_{12}\vert f_2>$ and
$g_{jj}^{-1}=E-\varepsilon_j$. (This representation is convenient,
because the self-energy matrix in Eq.~(\ref{GBB}) below is then
diagonal.)

The effect of coupling the left (right) dot to the left (right)
lead via a coupling matrix $W_1$ ($W_2$) is represented by self
energies $\Sigma_L = \sigma_L-i\Gamma_L$ ($\Sigma_R=
\sigma_R-i\Gamma_R$), (where $\sigma_L, \Gamma_L, \sigma_R,
\Gamma_R$ are real) defined by
\begin{equation}
\Sigma_L = \sum_{n_L}<f_1\vert W^\dagger_1\vert
n_L>g_A(n_L)<n_L\vert W_1\vert f_1>
\end{equation}
and
\begin{equation}
\Sigma_R = \sum_{n_R}<f_2\vert W^\dagger_2\vert
n_R>g_A(n_R)<n_R\vert W_2\vert f_2>,
\end{equation}
where $\vert n_{L\, (R)}>$ is a channel state  belonging to the L (R) lead
and $g_A(n_{L\,(R)})$ is the channel Green's function, such that 
\begin{equation} 
{g_A^{L \,(R)}=  \sum_{n_L\,(n_R)}\vert n_{L\,(R)} >
g_A(n_{L\,(R)}) < n_{L\, (R)}\vert}
\end{equation} 
is the corresponding surface Green's function.

 In the presence of the leads, the Green's function $G_{BB}$ of the double
 dot is given by
 \begin{equation}\label{GBB}
G_{DD}^{-1}=g_D^{-1} -\left(%
\begin{array}{cc}
  \Sigma_L & 0 \\
  0 & \Sigma_R \\
\end{array}%
\right),\\
\end{equation}
which yields the transmission coefficient $T_{dd}$ via the formula
\begin{equation}\label{tr-gammaGgammaG}
    T_{dd}=4{\rm Tr}[\Gamma(L)G_{DD}\Gamma(R)G^\dagger_{DD}]=4\Gamma_L \Gamma_R \vert (G_{DD})_{12}\vert^2,
\end{equation}
where
\begin{equation}
\Gamma(L)=\left(%
\begin{array}{cc}
  \Gamma_L & 0 \\
  0 & 0 \\
\end{array}%
\right),\\
\end{equation}
and
\begin{equation}
\Gamma(R)=\left(%
\begin{array}{cc}
  0 & 0 \\
  0 & \Gamma_R \\
\end{array}%
\right).\\
\end{equation}

Finally from Eq.~(\ref{GBB}),
\begin{equation}
(G_{DD})_{12}=
t_c/[(E-\varepsilon_1-\sigma_L+i\Gamma_L)(E-\varepsilon_2-\sigma_R+i\Gamma_R)
-\vert  t_c\vert^2],
\end{equation}
and writing $\tilde\varepsilon_1=\varepsilon_1+\sigma_L$ and
$\tilde\varepsilon_2=\varepsilon_2+\sigma_R$, yields equation (\ref{eq:transmission})
of the main text.

We note that this equation resembles Eq.~(20) of
Ref.~\onlinecite{ref:matveev2}. 
However the latter omits the self energy
terms $\sigma_L$ and $\sigma_R$, which in general are
non-negligible.

\section{}

\begin{figure}[htb]
\includegraphics[scale=0.2]{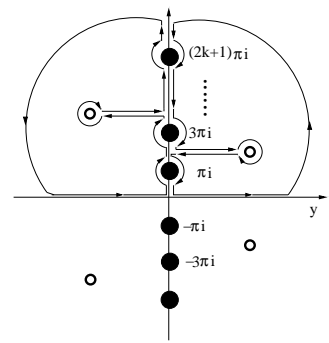}
\caption{The two integration contours to calculate the integral
in Eq.~(\ref{eq:finite_T_trans}). Filled circles denote the second order poles
of $-\frac{\partial f_0(E)}{\partial E}$, open circles show the poles of
$T_{dd}(E)$.
\label{fig:contours}}
\end{figure}

Introducing the dimensionless variable ${y=(E-E_F)/k_BT}$, the integral $I$ in Eq.~(\ref{eq:finite_T_trans}) reads
\begin{equation}
I=\frac{4\Gamma_L\Gamma_R\,t_c^2}{(k_B\,T)^4}\int_{-\infty}^{\infty}
\frac{\textrm{d}y f'(y)}{(y-\overline{\omega}_+)
(y-\overline{\omega}_+^{\,*})(y-\overline{\omega}_-)(y-\overline{\omega}_-^{\,*})}
\end{equation}
where
$f'(y)=1/\cosh^2(\frac{y}{2})$ and
 $\overline{\omega}_{\pm}=(E_{\pm}-E_F)/k_B\,T$. This integral can be
calculated using contour integration.
Care has to be taken, however,  because the integrand is not bounded on the imaginary
axes. Nevertheless, one can calculate this integral as a sum of two contour integrals,
as shown in Fig.~\ref{fig:contours}.

Closing the contours in the upper half plane, the
contributions of the two contours along the imaginary axes cancel, except
around  the  (second order) poles of the derivative of the Fermi function $f_0(E)$
(shown by filled circles in Fig.~\ref{fig:contours}).
These poles are located at $(2k+1)i\pi$, where $k$ is an integer.
The other contribution to the integral comes from the poles $E_{\pm}$ of the transmission
function $T_{dd}(E)$ (denoted by open circles in Fig.~\ref{fig:contours}).
Summing  all the contributions from the poles and
using the series expansions of the first polygamma function\cite{ref:abramowitz}
\begin{equation}
\psi^{(1)}(z)=\sum_{k=0}^{\infty}\frac{1}{(z+k)^2}
\end{equation}
and  of the $1/\cos^2(z)$ function
\begin{equation}
\frac{1}{\cos^2(z)}=4\sum_{k=0}^{\infty}\left[\frac{1}{((2k+1)\pi-z)^2}
+ \frac{1}{((2k+1)\pi+z)^2} \right],
\end{equation}
one can finally obtain the result shown in Eq.~(\ref{eq:polygammas}).

\end{document}